\documentclass[preprint,12pt]{elsarticle}

\usepackage{graphics}

\usepackage{amssymb}
\usepackage{latexsym}

\journal{Physica A}

\begin{document}

\begin{frontmatter}



\title{
Deconfined-critical behavior of the
VBS- and nematic-order parameters
for the spatially anisotropic $S=1$-spin model
}


\author{Yoshihiro Nishiyama} 

\address{Department of Physics, Faculty of Science,
Okayama University, Okayama 700-8530, Japan}

\begin{abstract}
The phase transition between
the valence-bond-solid (VBS) and nematic phases,
the so-called
deconfined criticality,
was
investigated for
the quantum $S=1$-spin model on the spatially anisotropic triangular
lattice
with the biquadratic interaction
by means of the numerical diagonalization method.
We calculated
both
VBS- and nematic-order parameters,
aiming to clarify the nature of this transition
from complementary viewpoints.
Simulating the clusters with $N \le 20$ spins,
we estimate the
correlation-length critical exponent
as $\nu=0.95(14)$.
We also calculated
Fisher's exponent (anomalous dimension) for each order parameter.
\end{abstract}

\begin{keyword}

75.10.Jm        Quantized spin models,
05.30.-d        Quantum statistical mechanics (for quantum fluids aspects, see 67.10.Fj),
75.40.Mg Numerical simulation studies,
74.25.Ha        Magnetic properties
\end{keyword}

\end{frontmatter}



\section{\label{section1}
Introduction}

According to the deconfined-criticality scenario 
\cite{Senthil04,Senthil04b,Levin04},
in two dimensions,
the phase transition
separating the 
valence-bond-solid (VBS) 
and antiferromagnetic phases
is continuous.
(Naively \cite{Senthil04b},
such a transition should be discontinuous,
because the adjacent phases possess distinctive
order parameters
such as the VBS-coverage pattern and the sublattice magnetization,
respectively.)
Extensive computer simulations have been made
to support this claim \cite{Sandvik07,Melko08}.
However, still,
one cannot exclude
a possibility of 
a weak-first-order transition accompanied with an appreciable latent-heat 
release \cite{Kotov09,Isaev10,Kotov10,Isaev10b,Kuklov04,Kuklov08,Jian08,Kruger06%
,Sirker06}.

The magnetic frustration is a good clue to the realization of
the VBS phase \cite{Sirker06,Poilblanc06,Nishiyama12}.
Alternatively, one is able to stabilize the VBS state
with the spatial anisotropy and the biquadratic 
interaction \cite{Harada07,Nishiyama09}.
In our preceding paper \cite{Nishiyama11},
we investigated
the quantum $S=1$-spin model
on the spatially anisotropic triangular lattice 
with the biquadratic interaction,
Eq. (\ref{Hamiltonian}),
and analyzed the singularity of the VBS-nematic phase transition;
a schematic phase diagram is presented 
in Fig. \ref{figure1},
where
the parameters $J$ and $j$ control
the
spatial anisotropy
and the biquadratic interaction, respectively.
Scrutinizing the scaling behavior of the
excitation gap, we obtained 
the correlation-length critical exponent
$\nu=0.92(10)$ \cite{Nishiyama11}.
In this paper,
we calculate both VBS- and nematic-order parameters,
and survey the criticality 
from complementary viewpoints.

To be specific, we present the Hamiltonian for
the $S=1$-spin model on the spatially anisotropic triangular lattice
with the biquadratic interaction
\begin{equation}
\label{Hamiltonian}
{\cal H}
= 
 -J \sum_{\langle ij \rangle } 
  [ j {\bf S}_i \cdot {\bf S}_j 
         + ({\bf S}_i \cdot {\bf S}_j)^2 ]
-
J' \sum_{\langle \langle ij \rangle\rangle} 
      ({\bf S}_i \cdot {\bf S}_j)^2 
.
\end{equation}
Here,
the quantum $S=1$ spins $\{ {\bf S}_i \}$ are placed at 
each triangular-lattice point
$i$; see Fig. \ref{figure2} (a).
The summation 
$\sum_{\langle ij \rangle}$ 
($\sum_{\langle \langle ij \rangle \rangle}$)
runs over all possible nearest-neighbor
(skew-diagonal) pairs.
The parameter $J(>0)$ ($J'$) 
denotes the corresponding coupling constant.
Hereafter, we consider $J'$ as the unit of energy ($J'=1$).
Along the $J$ bond, both quadratic and biquadratic interactions 
exist, and the parameter $j(>0)$ controls a strength of the former component.
The $J'$-bond interaction is purely biquadratic.
The interaction $J$ interpolates the one-dimensional ($J=0$)
and square-lattice ($J \to \infty$) structures;
correspondingly, there appear
the VBS \cite{Fath95} and 
spin-nematic \cite{Harada02} phases (Fig. \ref{figure1}).
In order to take into account such a geometrical character,
we implement the screw-boundary condition
[Fig. \ref{figure2} (b)] through resorting to Novotny's 
method \cite{Novotny92,Nishiyama08}
(Section \ref{section2_1}).

The critical indices for the deconfined criticality have been
investigated thoroughly.
We overview a number of related studies.
First,
the
$S=1$-spin model on the spatially anisotropic square 
lattice
with the biquadratic interaction \cite{Harada07} 
was simulated by means of the quantum Monte Carlo method;
here, the spatial-anisotropy axis is set to be parallel to
the primitive vector of the unit cell.
At the VBS-nematic phase boundary,
the authors estimated the (reciprocal) correlation-length
critical exponent as 
$y_V[=(\nu^V)^{-1}]=2.8(2)$ 
and 
$y_Q[=(\nu^Q)^{-1}]=2.8(2)$
for the VBS- and
nematic-order parameters, respectively.
These indices coincide, as anticipated; namely, 
a relation $\nu=\nu_V=\nu_Q$ holds.
Additionally, they obtained Fisher's exponent for the nematic order
$\eta_Q=1.25(20)-z$;
postulating the dynamical critical exponent $z=1$ (see below),
one arrives at $\eta_Q=0.25(20)$.
(Its VBS-order counterpart $\eta_V$ was not given.)
In our model, Eq. (\ref{Hamiltonian}),
we impose
the spatial anisotropy along the skew-diagonal
direction [Fig. \ref{figure2} (a)], 
aiming to make an alternative approach to this problem.
Second, the internal symmetry of each spin
is extended from an ordinary SU$(2)$ to
SU$(N)$ with an arbitrary integral number $N$
\cite{Read89,Kawashima07,Lou09,Kaul11}
or even a continuously variable parameter $N$
\cite{Beach09};
note that the $S=1$-spin model with the (finely-tuned)
biquadratic interaction can possess
the SU$(3)$ symmetry, and 
this problem is quite relevant to ours.
The latter model \cite{Beach09} demonstrates a clear evidence of a 
deconfined
criticality 
with the critical indices,
$0.75 < \nu < 1$, $z=1$, and  $\eta=0.63(4)$ at $N_c=4.57(5)$;
here, the index $\eta$ denotes the ordinary Fisher's exponent 
for the constituent SU$(N)$ moment.
It would be noteworthy that Fisher's exponent
acquires a considerable enhancement.
Third,
pioneering considerations \cite{Sandvik07,Melko08} on the deconfined criticality
were set forward for
the 
$S=1/2$ square-lattice model with the biquadratic 
(plaquette-four-spin) interaction.  
Sets of critical indices,
$\nu=0.78(3)$, $\eta=\eta_V=0.26(3)$, and 
$z=1$ \cite{Sandvik07};
$\nu=0.68(4)$, $\eta=0.35(3)$,    
and $z=1$
\cite{Melko08}, have been reported.
A closely related model, the so-called $Q_v$ model \cite{Lou09b},
yields almost similar results,
$\nu=0.78(3)$, $\beta=0.27(2)$, and $\beta_V=0.68(3)$;
here, the indices, $\beta$ and $\beta_V$, are the 
magnetization critical
exponents
for the constituent moment
and the VBS-order parameter, respectively.  
Last, we recollect a number of field-theoretical considerations.
A formalism for the $S=1$-spin model
is provided in Ref. \cite{Grover11},
where a microscopic origin of an enhancement of Fisher's exponent
is argued. 
An extensive simulation on the hedgehog-suppressed O$(3)$ model \cite{Motruich04}
revealed a clear evidence for the novel critical indices, $\nu=1.0(2)$ and $\eta=0.6(1)$.
As a reference, 
we quote the critical exponents for
the
three-dimensional
Heisenberg universality 
class \cite{Compostrini02},
$\nu=0.7112(5)$ and $\eta=0.0375(5)$. 

The rest of this paper is organized as follows.
In Section \ref{section2}, we present the simulation results.
Technical details are 
explained as well.
In Section \ref{section3}, we address the summary and discussions.

\section{\label{section2}
Numerical results}

In this section, we
present the simulation results.
Before commencing detailed analyses of criticality,
we 
explain the simulation algorithm in a self-contained manner.
Throughout this section,
we fix
the parameter $j$ to 
an intermediate value,              
$j/J'=0.5$, where the finite-size-scaling behavior improves \cite{Nishiyama11};
see the phase diagram, Fig. \ref{figure1}. 
The number of spins (system size) ranges in
$N=10,12,\dots,20$.
The linear dimension $L$ of the cluster is
given by 
\begin{equation}
L=\sqrt{N}
  ,
\end{equation}
because $N$ spins form a rectangular cluster.

The Hamiltonian (\ref{Hamiltonian}) possesses a number of symmetries,
with which
one is able to reduce (block-diagonalize)
the size of the Hamiltonian matrix.
Here, aiming to eliminate the Hilbert-space dimensionality,
we look into
the subspace with the total longitudinal-spin moment
$\sum_{i=1}^N S^z_i=0$,
even parity,
and the wave number $k=0$ 
with respect to the internal-spin-rotation-,
spin-inversion-
($S^z_i \to -S^z_i$), 
and
lattice-translation-symmetry groups, respectively.
The ground state belongs to this subspace.
The size of the reduced (block-diagonal) Hamiltonian
is $9436203$ for $N=20$.


\subsection{\label{section2_1}
Simulation method
}

As mentioned in the Introduction,
we impose the screw-boundary condition on a finite cluster
with $N$ spins; see Fig. \ref{figure2} (b).
Basically, the spins, 
$\{ {\bf S}_i \}$ ($i \le N$),
 constitute a one-dimensional ($d=1$) structure,
and the dimensionality is lifted to $d=2$ by the bridges over
the long-range pairs.
According to Novotny \cite{Novotny92},
 the long-range interactions
are
introduced systematically
by the use of the
translation operator $P$; see Eq. (\ref{HXXX}), for instance.
The operator $P$ satisfies the formula
\begin{equation}
P | S_1,S_2,\dots,S_N \rangle
   = | S_N,S_1,\dots,S_{N-1}\rangle  .
\end{equation}
Here, the base $|\{ S_i\}\rangle$ diagonalizes each of $\{ S^z_i\}$;
namely, the relation
$S^z_k | \{S_i\}\rangle = S_k | \{ S_i\}\rangle  $
holds.
Novotny's method was adapted to the quantum $S=1$ $XY$ model
in $d=2$ dimensions \cite{Nishiyama08}.
Our simulation scheme is based on this formalism.
In the following, 
we present the modifications explicitly for the sake of self-consistency.
The $XY$ interaction $H_{XY}$, Eq. (4) of Ref. \cite{Nishiyama08},
has to be replaced with the Heisenberg interaction
\begin{equation}
\label{HXXX}
H_{XXX}(v)=\sum_{i=1}^{N}(
P^vS^x_iP^{-v}S^x_i+
P^vS^y_iP^{-v}S^y_i+
P^vS^z_iP^{-v}S^z_i)    
  .
\end{equation}
Additionally, we introduce the biquadratic interaction
\begin{equation}
H_4(v)=-\frac{1}{2} H_{XXX}(v)
+\frac{1}{2} \sum_{i=1}^{N}
   \sum_{\alpha=1}^5
P^v Q^\alpha_i P^{-v} Q^\alpha_i
.
\end{equation}
Here, we utilized an equality
\begin{equation}
({\bf S}_i \cdot {\bf S}_j)^2 =
  -{\bf S}_i \cdot {\bf S}_j/2
  +\sum_{\alpha=1}^{5} Q_i^\alpha Q_j^\alpha/2+4/3
,
\end{equation}
with the quadrapole moments,
$Q_i^1 = (S_i^x)^2-(S_i^y)^2$,
$Q_i^2 = [2(S_i^z)^2-(S_i^x)^2-(S_i^y)^2]/ \sqrt{3}$,
$Q_i^3 = S_i^xS_i^y+S_i^yS_i^x$,
$Q_i^4 = S_i^yS_i^z+S_i^zS_i^y$,
and
$Q_i^5 = S_i^xS_i^z+S_i^zS_i^x$.
Based on these expressions,
we replace Eq. (3) of Ref. \cite{Nishiyama08} with
\begin{equation}
\label{Novotny_Hamiltonian}
{\cal H}
=
-J[j H_{XXX}(\sqrt{N})+j H_{XXX}(\sqrt{N}-1)
     +H_4(\sqrt{N})+H_4(\sqrt{N}-1)]
-J' H_4(1)
.
\end{equation}
We diagonalize this matrix for $N \le 20$ spins. 
The above formulae complete the formal basis of our simulation
scheme.
However,
in order to evaluate the above Hamiltonian-matrix elements
efficiently,
one may refer to a number of techniques addressed in Refs. \cite{Novotny92,Nishiyama08}.

\subsection{\label{section2_2}
Critical behavior of the VBS-order parameter}

In this section, we investigate
the critical behavior of the VBS-order parameter;
in the VBS phase, along the $J'$-bond direction,
the staggered-dimer order develops \cite{Fath95}.

In Fig. \ref{figure3},
we present
the Binder parameter
\begin{equation}
\label{Binder_V}
U_V =1- 
  \frac{\langle m_V^4 \rangle}
       {3\langle m_V^2 \rangle^2} 
     ,
\end{equation}
for various $J(/J')$ and $N=10,12,\dots,20$.
Here, the quadrature of the VBS-order parameter 
is given by
\begin{equation}
\label{magnetization_V}
m_V^2 = \frac{
S^z_1 S^z_2
             }{N}
\sum_{i=2}^{N-1}  (-1)^i 
          S^z_{1+i} S^z_{2+i}  ,
\end{equation}
and
the symbol $\langle \cdots \rangle$ denotes the 
ground-state average. 
The interaction parameter $j$ is set to an intermediate
value $j=0.5$, as mentioned above.
As the system size increases,
the Binder parameter increases (decreases)
in the long- (short-) range-order phase.
Hence, in Fig. \ref{figure3}, we see that
the VBS order develops in the small-$J$ regime;
low-dimensionality promotes the formation of the VBS state.
The intersection point of the curves indicates a location of the critical point.

In order to extrapolate the critical (intersection)
point to the thermodynamic limit,
in Fig. \ref{figure4},
we plot the approximate critical point
$J^V_c(N_1,N_2)$  (pluses)
for $[2/(N_1+N_2)]^2$ with $10 \le N_1<N_2 \le 20$;
the parameters are the same as those of Fig. \ref{figure3}.
Here, the approximate transition point,
$J_c^V(N_1,N_2)$,
 denotes a scale-invariant point with
respect to a pair of system sizes $(N_1,N_2)$.
Namely, the relation 
\begin{equation}
\label{critical_point}
 U_\alpha (N_1) |_{J=J^\alpha_c(N_1,N_2)} = U_\alpha (N_2) |_{J=J^\alpha_c(N_1,N_2)}
  ,  
\end{equation}
with $\alpha=V$
holds.
The least-squares fit to the data of Fig. \ref{figure4}
yields an estimate $J_c^V=0.236(18)$ in the thermodynamic limit,
$N \to \infty$.
A remark is in order.
The wavy character in Fig. \ref{figure4}
is an artifact due to the screw-boundary condition.
Actually, there appears
a bump (drop) at $N\approx 4^2$  
($\approx 3^2$, $5^2$); in other words,
an undulation comes out,
depending on the condition whether the linear dimension of the cluster,
$\sqrt{N}$,
is close to an even number ($N \approx 4^2$) or an odd one 
($N \approx 3^2$, $5^2$).
Possible systematic error of $J_c^V$
is argued in Section \ref{section2_5}.

We turn to the analysis of the correlation-length critical exponent
$\nu$.
In Fig. \ref{figure5},
we plot the approximate critical exponent (pluses)
\begin{equation}
\label{critical_exponent_nu}
\nu^{\alpha} (N_1,N_2)=
\frac{ \ln(L_1/L_2) }{
\ln [ \partial_J U_\alpha(N_1) /\partial_J U_\alpha(N_2) ] |_{J=J_c^\alpha(N_1,N_2)}  }
           ,
\end{equation}
for $[2/(N_1+N_2)]^2$ with $\alpha=V$ and 
$10 \le N_1<N_2 \le 20$ ($L_{1,2}=\sqrt{N_{1,2}}$).
The parameters are the same as those of Fig. \ref{figure3}.
Again, there emerges a wavy character intrinsic to the screw-boundary
condition; a notable bump at $N \approx 4^2$ would be an artifact,
preventing us to access to the thermodynamic limit systematically.
The least-squares fit to these data yields $\nu^V=1.001(93)$
in the thermodynamic limit.
Uncertainty of this result is argued 
in 
Section \ref{section2_5}.

\subsection{\label{section2_3}
Critical behavior of the nematic- (quadrupolar-) order parameter}

In this section, we investigate
the critical behavior of
the nematic- (quadrupolar-) order parameter.

In Fig. \ref{figure6},
we present 
the Binder parameter
\begin{equation}
\label{Binder_Q}
U_Q =1- 
  \frac{\langle m_Q^4 \rangle}
       {3\langle m_Q^2 \rangle^2} ,
\end{equation}
for various $J$ and $N=10,12,\dots,20$.
Here, 
the nematic-order parameter 
is given by
\begin{equation}
\label{magnetization_Q}
m_Q^2 =  \sum_{i=N/2}^{N/2+1} Q^1_1 Q^1_{1+i}
   .
\end{equation}
The parameters are the same as those of Fig. \ref{figure3}.
The nematic order appears to develop in the
large-$J$ regime,
offering a sharp contrast to the VBS order (Fig. \ref{figure3}).
We amended the definition of $m^2_Q$,
Eq. (\ref{magnetization_Q}),
so as to attain an intersection point of the $U_Q$ curves in Fig. \ref{figure6};
otherwise, the intersection point disappears. 
Namely,
we discarded
the short-distance contributions within the $m^2_Q$ correlator
in order to get rid of corrections to scaling.
As a byproduct, the location of intersection point
drifts significantly with respect to $N$;
a subtlety of $Q$-based result 
is reconsidered in Section \ref{section2_5}.

In Fig. \ref{figure4},
we plot the approximate transition point
$J^Q_c(N_1,N_2)$  (crosses), Eq. (\ref{critical_point}),
for $[2/(N_1+N_2)]^2$ with $10 \le N_1<N_2 \le 20$.
The parameters are the same as those of Fig. \ref{figure3}.
The finite-size errors seem to be larger than those of $J^V_c$.
The least-squares fit to the data of Fig. \ref{figure4}
yields an estimate $J^Q_c= 0.151(11)$ in the thermodynamic limit;
the error margin is considered
in Section \ref{section2_5}

We turn to the analysis of the correlation-length critical exponent
$\nu^Q(N_1,N_2)$, Eq. (\ref{critical_exponent_nu}).
In Fig. \ref{figure5},
we plot the approximate critical exponent
$\nu^Q(N_1,N_2)$ (crosses) 
for $[2/(N_1+N_2)]^2$ with 
$10 \le N_1<N_2 \le 20$.
The parameters are the same as those of Fig. \ref{figure3}.
The least-squares fit to these data yields $\nu^Q=0.902(45)$
in the thermodynamic limit;
a possible systematic error is appreciated 
in Section \ref{section2_5}.

\subsection{\label{section2_4}
Fisher's critical exponent $\eta_{V,Q}$}

At the critical point $J=J_c$, 
the quadratic moment $m_{V,Q}^2$
obeys the power law,
$m^2_{V,Q} \sim 1/L^{1+\eta_{V,Q}}$,
with
Fisher's exponent (anomalous dimension)
$\eta_{V,Q}$ and system size $L$.

First, we consider the case of $m_V^2$.
In Fig. \ref{figure7},
we plot the approximate critical exponent
\begin{equation}
\label{critical_exponent_etaV}
\eta_V^\alpha(N_1,N_2)=  -
\frac{
\ln [  m^2_V(N_1)/m^2_V(N_2) ] |_{J=J_c^\alpha(N_1,N_2)}  }
{ \ln(L_1/L_2) }
-1
           ,
\end{equation}
for $[2/(N_1+N_2)]^2$, $\alpha=V$ (plusses), and $\alpha=Q$ (crosses)
with
$10 \le N_1<N_2 \le 20$.
The parameters are the same as those of Fig. \ref{figure3}.
The least-squares fit to these data yields $\eta_V^V=0.786(38)$
and $\eta_V^Q=-0.12(16)$
in the thermodynamic limit.

Second, we consider Fisher's exponent for the nematic order.
In Fig. \ref{figure8},
we plot the approximate critical exponent
\begin{equation}
\label{critical_exponent_etaQ}
\eta_Q^\alpha(N_1,N_2)=  -
\frac{
\ln [ m^2_Q(N_1)/m^2_Q(N_2) ] |_{J=J_c^\alpha(N_1,N_2)}  }
{ \ln(L_1/L_2) }
-1
           ,
\end{equation}
for $[2/(N_1+N_2)]^2$, $\alpha=V$ (plusses), and $\alpha=Q$ (crosses)
with
$10 \le N_1<N_2 \le 20$.
The parameters are the same as those of Fig. \ref{figure3}.
The least-squares fit to these data yields
 $\eta_Q^V=-0.141(94)$ and $\eta_Q^Q=-0.089(18)$
in the thermodynamic limit.

\subsection{\label{section2_5}
Extrapolation errors of $J_c$, $\nu$ and $\eta_{V,Q}$
}

In this section,
we consider possible systematic errors of
$J_c$, $\nu$, $\eta_{V}$, and $\eta_Q$
obtained in
Figs. \ref{figure4}, \ref{figure5}, \ref{figure7}, and \ref{figure8},
respectively.

First, we consider the critical point $J_c(=J_c^{V,Q})$.
In Fig. \ref{figure4}, we made
independent extrapolations, which yield
$J_c^V=0.236(18)$ and $J_c^Q=0.151(11)$.
The discrepancy (systematic error) between these results,
$\approx 0.09$, is larger than the
insystematic (statistical) errors, O($ 10^{-2}$).
Hence, we consider the former as the main source of the extrapolation error.
Taking a mean value of $J_c^V$ and $J_c^Q$,
we estimate $J_c=0.19(9)$.
We address a number of remarks.
First, this estimate is consistent with the preceding one
$J_c =0.285(5)$ \cite{Nishiyama11} within the error margin;
note that the preceding estimate \cite{Nishiyama11}
was calculated through the scaling of the excitation gap. 
Second, the result $J_c^V$ would be more reliable than $J_c^Q$;
actually, the slope of $J_c^V(N_1,N_2)$ in Fig. \ref{figure4}
appears to be smaller than that of $J_c^Q$, suggesting that
the former extrapolation would be trustworthy.
Last, the extrapolated critical point $J_c$ does not affect the subsequent analyses;
rather,
the approximate critical point
$J_c^\alpha(N_1,N_2)$ was fed into the
formulas for the critical indices, 
Eqs. 
(\ref{critical_exponent_nu}),
(\ref{critical_exponent_etaV}), and
(\ref{critical_exponent_etaQ}).

Second, we turn to 
the correlation-length critical exponent $\nu(=\nu^{V,Q})$.
In Fig. \ref{figure5},
two independent extrapolations yield
$\nu^V=1.001(93)$ and $\nu^Q=0.902(45)$.
The discrepancy,
$\sim 0.1$, is comparable with 
the statistical
error, 
$\sim 0.1$.
Taking a mean value, we obtain an estimate
\begin{equation}
\label{result_nu}
\nu=0.95(14)    .
\end{equation}
Here,
the error margin comes from 
$0.14(\approx\sqrt{0.1^2+0.1^2})$ 
(propagation of uncertainty) with
systematic ($\sim 0.1$) and
insystematic ($\sim 0.1$) errors.
The estimate, Eq. (\ref{result_nu}), agrees with the
preceding one $\nu=0.92(10)$ \cite{Nishiyama11}.
Our result may support the deconfined-criticality scenario,
suggesting that the exponent $\nu$ acquires an enhancement,
as compared with 
that of
the three-dimensional Heisenberg universality class,
$\nu=0.7112(5)$ 
\cite{Compostrini02}.

Last, we consider Fisher's exponent $\eta_{V,Q}$.
In Fig. \ref{figure7}, 
the results,
$\eta_V^V=0.786(38)$
and 
$\eta_V^Q=-0.12(16)$, are obtained.
The discrepancy between them, $\sim 0.9$,
dominates the statistical error, $\sim 0.03$.
Considering the former as a error margin,
we obtain $\eta_V=0.3(9)$.
On the one hand,
in Fig. \ref{figure8},  the discrepancy
between
$\eta_Q^V=-0.141(94)$ and
$\eta_Q^Q=-0.089(18)$ is negligible.
Considering the statistical error $\sim 0.1$ as a main source of uncertainty,
we arrive at $\eta_Q=-0.1(1)$.

This is a good position to address a remark.
As mentioned above,
the VBS-order-based results,
 $J^V_c$ and $\eta^V_{V,Q}$,
are more reliable than the nematic-order-based ones,
$J^Q_c$ and $\eta^Q_{V,Q}$.
In general, the discrete symmetry, $m^2_V$,
is more robust than the continuous one,
$m^2_Q$,
allowing us to make a systematic scaling analysis
even for the system size tractable with the numerical
diagonalization method.
In fact,
as claimed in Section III F 2  of Ref. \cite{Manousakis91},
the numerical diagonalization method
up to $N=4\times4$ is incapable of providing a conclusive evidence for
the spontaneous magnetization of the two-dimensional Heisenberg antiferromagnet.
Hence,
tentatively, we discard
the nematic-order-based results,
and
refer to the VBS-order-based ones
to obtain crude estimates,
$\eta_V\approx 0.8$ and $\eta_Q\approx -0.1$.
These conclusions are comparable with 
the large-scale-Monte-Carlo results \cite{Lou09b},
$\eta_V=0.74(10)$ and $\eta=-0.31(6)$,
for the $Q_v$ model;
here, we made use of the scaling relation, $1+\eta=2\beta/\nu$,   
to evaluate $\eta_V$ and $\eta$
from $(\beta_V,\beta)=[0.68(3),0.27(2)]$ and $\nu=0.78(3)$
\cite{Lou09b}.
Possibly \cite{Lou09b},
the exponent $\eta$ suffers from ``drift'' (scaling corrections),
and may restore $\eta >0$ through taking into account of yet unidentified
scaling corrections.
As mentioned in the Introduction,
in the preceeding study of the $S=1$-spin model
\cite{Harada07},
an estimate
$\eta_Q=0.25(20)$
was reported.
It is suggested that 
the anomalous dimension for the nematic order
$\eta_Q$ would be almost vanishing.

\section{\label{section3}
Summary and discussions}

\begin{figure}
\includegraphics{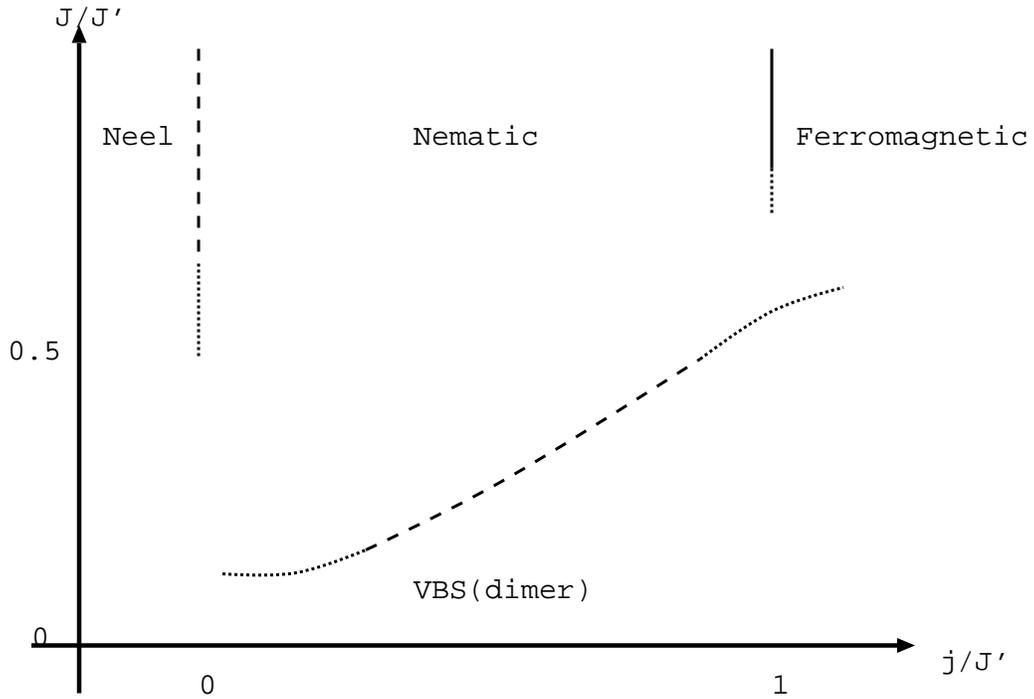}%
\caption{  \label{figure1}
A schematic phase diagram for the 
$S=1$-spin model
on the 
spatially anisotropic triangular lattice
with the biquadratic interaction, Eq. 
(\ref{Hamiltonian}),
is presented.
The limiting cases $J = 0$ and $ J \to \infty$
were studied in Refs. \cite{Fath95} and \cite{Harada02},
respectively.
The solid (dashed) lines stand for the phase boundaries
of discontinuous (continuous) character.
The dotted lines are ambiguous.
We investigate the phase boundary separating the nematic and
VBS phases.
}
\end{figure}

\begin{figure}
\includegraphics{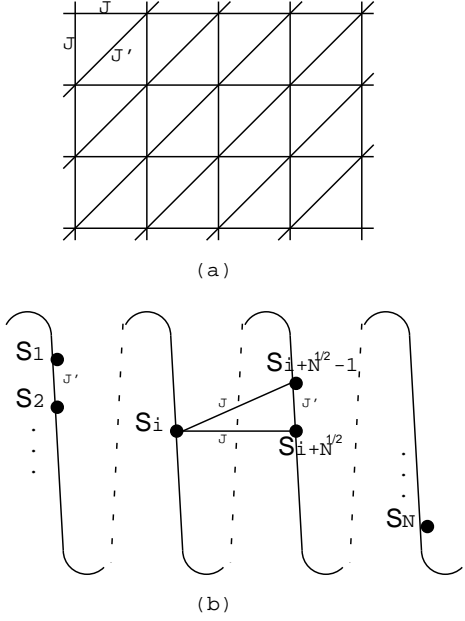}%
\caption{  \label{figure2}
(a)
We consider a spatially anisotropic triangular lattice;
the Hamiltonian is given by Eq. (\ref{Hamiltonian}).
The interaction $J$ interpolate the one- 
and two-dimensional 
lattice structures in the limiting cases of $J=0$ and $J\to\infty$,
respectively.
(b)
In order to take into account such a geometrical character,
we implement the screw-boundary condition.
As shown in the drawing,
a basic structure of the cluster is
an alignment of spins $ \{ {\bf S}_i \}$
($i \le N$).
Thereby,
the dimensionality is lifted to 
$d=2$ by the bridges over the ($\sqrt{N}$)-th neighbor
pairs through the $J$ bonds.
Technical details are 
explained in Section \ref{section2_1}.
}
\end{figure}

\begin{figure}
\includegraphics{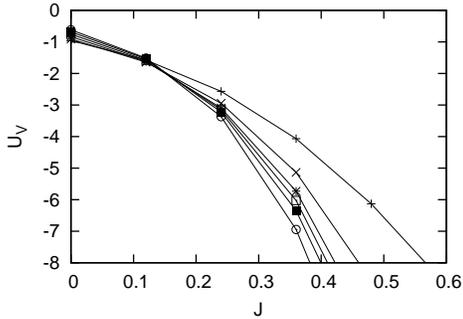}%
\caption{  \label{figure3}
The Binder parameter for the VBS-order parameter $U_V$
(\ref{Binder_V})
is plotted for various $J$ and $N=$
($+$) $10$,
($\times$) $12$,
($*$) $14$,
($\Box$) $16$,
($\blacksquare$) $18$, and
($\circ$) $20$.
The parameter $j$ is set to $j=0.5$.
($J'$ is the unit of energy.)
The VBS order develops in the small-$J$ side.
}
\end{figure}

\begin{figure}
\includegraphics{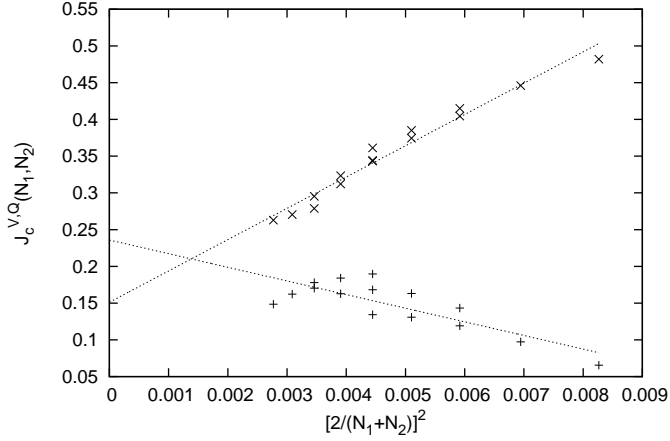}%
\caption{  \label{figure4}
The approximate critical point 
$J_c^\alpha (N_1,N_2)$
(\ref{critical_point})
is plotted for $[2/(N_1+N_2)]^2$ ($10 \le N_1 < N_2 \le 20$) 
with
$\alpha=V$ (plusses) and 
$\alpha=Q$ (crosses).
The parameters are the same as those of Fig. \ref{figure3}.
The least-squares fit to these data yields 
$J_c^V=0.236(18)$ 
and $J_c^Q=0.151(11)$, respectively,
in the thermodynamic limit.
}
\end{figure}

\begin{figure}
\includegraphics{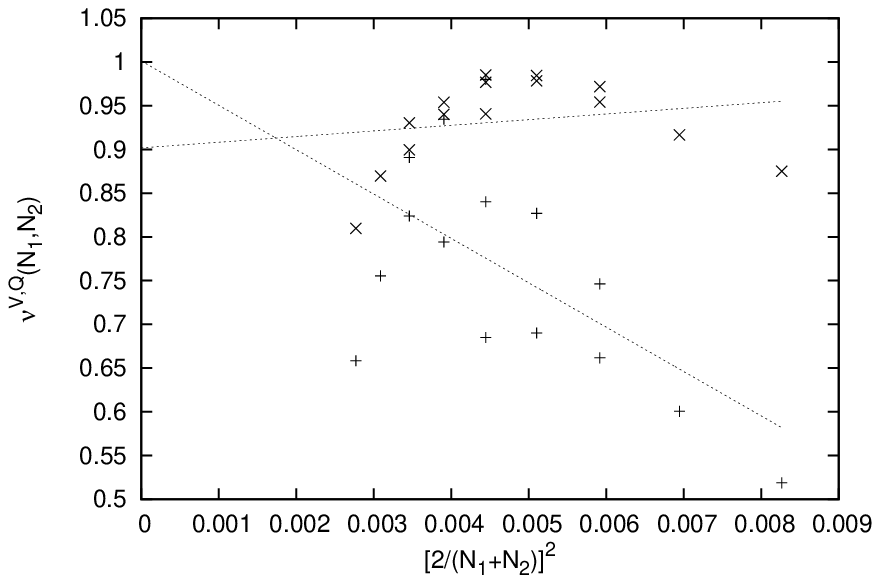}%
\caption{  \label{figure5}
The approximate critical exponent $\nu^\alpha(N_1,N_2)$
(\ref{critical_exponent_nu})
is plotted for $[2/(N_1+N_2)]^2$ ($10 \le N_1 < N_2 \le 20$)
with
$\alpha=V$ (plusses) and 
$\alpha=Q$ (crosses).
The parameters are the same as those of Fig. \ref{figure3}.
The least-squares fit to these data yields 
$\nu^V=1.001(93)$
and $\nu^Q=0.902(45)$, respectively,
in the thermodynamic limit.
}
\end{figure}

\begin{figure}
\includegraphics{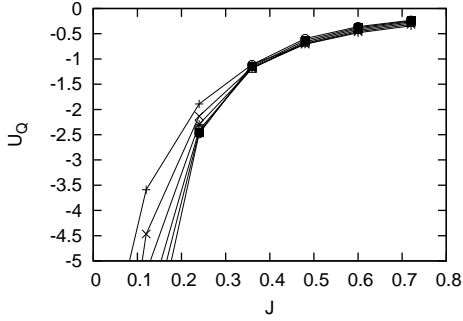}%
\caption{  \label{figure6}
The Binder parameter for the nematic-
(quadrupolar-) order parameter $U_Q$
(\ref{Binder_Q})
is plotted for various $J$ and $N=$
($+$) $10$,
($\times$) $12$,
($*$) $14$,
($\Box$) $16$,
($\blacksquare$) $18$, and
($\circ$) $20$.
The parameters are the same as those of Fig. \ref{figure3}.
The nematic order develops in the large-$J$ side.
}
\end{figure}

\begin{figure}
\includegraphics{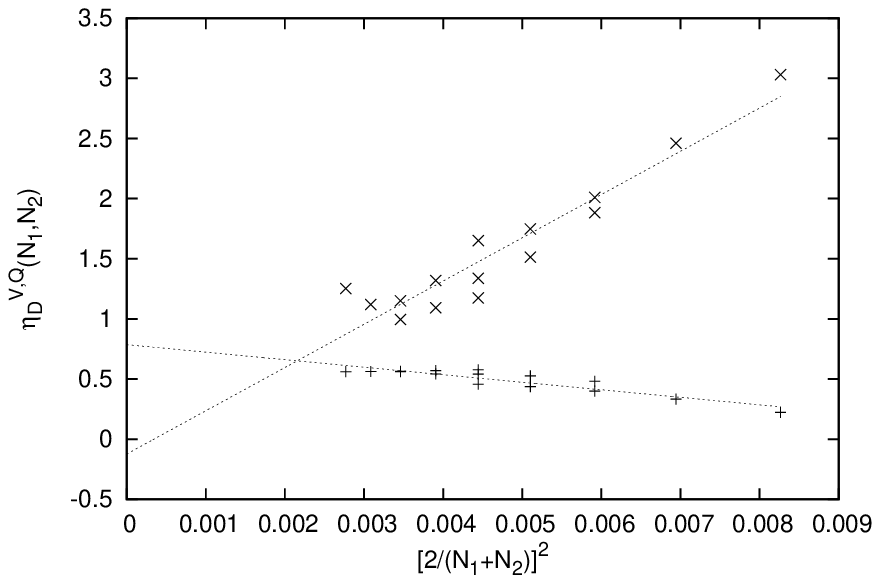}%
\caption{\label{figure7}
The approximate critical exponent $\eta_V^\alpha(N_1,N_2)$
(\ref{critical_exponent_etaV})
is plotted for $[2/(N_1+N_2)]^2$ ($10 \le N_1 < N_2 \le 20$)
with
$\alpha=V$ (pluses) and $\alpha=Q$ (crosses).
The parameters are the same as those of Fig. \ref{figure3}.
The least-squares fit to these data yields 
$\eta_V^V=0.786(38)$
and $\eta^Q_V=-0.12(16)$, respectively,
in the thermodynamic limit.
}
\end{figure}

\begin{figure}
\includegraphics{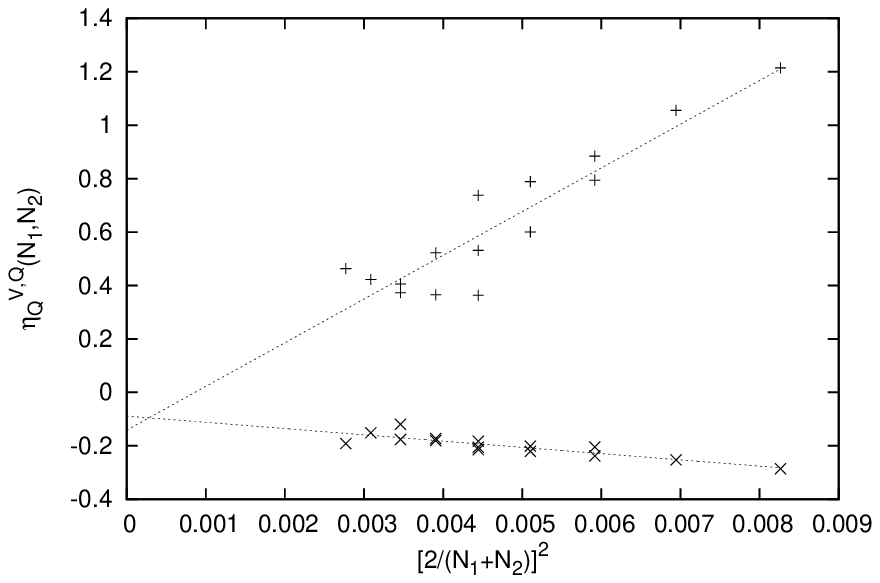}%
\caption{\label{figure8}
The approximate critical exponent $\eta_Q^\alpha(N_1,N_2)$
(\ref{critical_exponent_etaQ})
is plotted for $[2/(N_1+N_2)]^2$ ($10 \le N_1 < N_2 \le 20$)
with
$\alpha=V$ (pluses) and 
$\alpha=Q$ (crosses).
The parameters are the same as those of Fig. \ref{figure3}.
The least-squares fit to these data yields 
$\eta^V_Q=-0.141(94)$
and $\eta_Q^Q=-0.089(18)$, respectively,
in the thermodynamic limit.
}
\end{figure}

The phase transition separating the VBS and nematic phases (Fig. \ref{figure1}),
the so-called deconfined criticality,
was investigated for
the $S=1$-spin model on the spatially anisotropic triangular lattice
with the biquadratic interaction, Eq. (\ref{Hamiltonian}).
So far,
the criticality has been analyzed via the scaling of the first excitation
gap \cite{Nishiyama11}.
In this paper, evaluating both VBS- and nematic-order parameters,
we made complementary approaches to this criticality.
As a result, we estimate the 
correlation-length critical exponent $\nu=0.95(14)(\approx \nu^{V,Q})$; 
this estimate agrees with $\nu=0.92(10)$ \cite{Nishiyama11},
supporting the deconfined-criticality scenario.

Encouraged by this finding,
we put forward the analysis of
Fisher's exponent
$\eta_{V,Q}^\alpha$ for each order parameter.
As overviewed in the Introduction,
so far, the exponent $\eta_Q=0.25(20)$ \cite{Harada07}
has been reported as to the $S=1$-spin model;
corrections to scaling did not admit to the estimation of $\eta_V$.
We obtained a set of indices $(\eta_V,\eta_Q)=[0.3(9),-0.1(1)]$;
again, the exponent $\eta_V$ suffers from scaling corrections.
Alternatively,
setting $\alpha=V$ tentatively (Sec. \ref{section2_5}),
we arrive at crude results 
$(\eta_V,\eta_Q) \sim (0.8,-0.1)$.  
These results might be reminiscent of
the large-scale-simulation results \cite{Lou09b},
$\nu=0.78(3)$, 
$\eta_V=0.74(10)$, and $\eta=-0.31(6)$, for the $Q_v$ model.   
We suspect that the exponent $\eta_Q$ would almost vanish
through taking into account of corrections to scaling properly.
In Ref. \cite{Banerjee10}, Fisher's exponent was estimated rather
accurately
through scrutinizing
the local-moment distribution around a magnetic impurity;
this idea has a potential applicability to a wide class of systems.
This problem will be addressed in future study.

\bibliographystyle{elsarticle-num}







\end{document}